\documentclass[12pt]{iopart} 
\usepackage{graphicx}
\usepackage{dcolumn}
\usepackage{bm}
\usepackage{epsfig}
\usepackage{color}
\usepackage{psfrag}
\usepackage{stackrel}
\usepackage{amsfonts}

\newcommand{\be}{\begin{equation}}
\newcommand{\ee}{\end{equation}}
\newcommand{\ba}{\begin{eqnarray}}
\newcommand{\ea}{\end{eqnarray}}

\begin{document}
\title[Potts critical frontiers of bow-tie lattices]{Potts critical frontiers of inhomogeneous and asymmetric bow-tie lattices}

\author{Christian R.\ Scullard$^{1}$ and Jesper Lykke Jacobsen$^{2,3}$}
\address{${}^1$Livermore, California, USA}
\address{${}^2$LPTENS, \'Ecole Normale Sup\'erieure, 24 rue Lhomond, 75231
Paris, France}
\address{${}^3$Universit\'e Pierre et Marie Curie, 4 place Jussieu, 75252 Paris,
France}

\eads{\mailto{scullard1@llnl.gov}, \mailto{jesper.jacobsen@ens.fr}}



\begin{abstract}
We study the critical frontiers of the Potts model on two-dimensional bow-tie lattices with fully inhomogeneous coupling constants.
Generally, for the Potts critical frontier to be found exactly, the underlying lattice must be a 3-uniform hypergraph. A more general class of lattices 
are the 4-uniform ones, with unit cells contained within four boundary vertices.
We demonstrate that in some cases, such lattices can be decomposed into triangular cells, and solved using a modification of standard techniques. This leads to the exact inhomogeneous Potts critical frontiers on various lattices, such as the bow-tie lattice with five different couplings, and critical points for asymmetric bow-tie lattices.
\end{abstract}

\noindent

\section{Introduction}
Since its introduction sixty years ago \cite{Potts1952,Wu1982}, the Potts model has been the subject of intense study. Beneath its simple definition lies a wealth of interesting mathematical problems, many of which remain unsolved. One of these is the determination of the critical temperature, or more generally the critical frontier, for the transition to the disordered phase. Although the problem is solved in one dimension ($d=1$), it is completely unsolved for $d \geq 3$, and in two dimensions, critical temperatures can be found exactly only on a certain class of periodic lattices \cite{Wu2006}. Here, we aim to extend this class.

The $q$-state Potts model is defined as follows. Consider a $d$-dimensional lattice on which we assign to each vertex, $i$, a spin, $\sigma_i$, which can be in any of the discrete states $(1,...,q)$. The Hamiltonian for the system is defined as
\begin{equation}
-\beta H = K \sum_{\langle ij \rangle} \delta(\sigma_i,\sigma_j) \ , \label{eq:H}
\end{equation}
where $\beta = 1/(k_B T)$, $K$ is the coupling constant and the sum is over all nearest-neighbors, $\langle ij \rangle$. The corresponding partition function is given by
\begin{equation}
Z = \sum_{\{ \sigma \}} e^{-\beta H} \,, \label{eq:partZ}
\end{equation}
where $\{ \sigma \}$ indicates that the summation is over all possible spin configurations.

When $K>0$, it is energetically favorable for neighboring spins to align and we are in the ferromagnetic regime of the model. In particular, a ground state ($T=0$ or $K \rightarrow + \infty$) is one in which all spins are in the same state. Obviously, there are precisely $q$ such ground states. When $K<0$, we are in the {\it antiferromagnetic} regime, and the tendency of spins is to anti-align. A ground state ($K \rightarrow -\infty$) of such a system is now one in which neighboring spins avoid alignment as far as the lattice connectivity allows. In some cases this competition between anti-alignment and lattice connectivity leads to an extensive entropy of the ground state ensemble.

The formulation of the Potts model given by (\ref{eq:H}) and (\ref{eq:partZ}) is useful for its connection to a physical process. However, it is often more convenient to employ a cluster representation. Let us denote
the graph (lattice) on which the model is defined by $G=(V,E)$, where $V$ is the set of vertices and $E = \langle ij \rangle$ is the set of edges. Using the temperature parameter $v=e^K-1$ and summing (\ref{eq:partZ}) over the spins, the partition function on $G$ can be written \cite{FK72}
\begin{equation}
 Z= \sum_{E' \subseteq E} v^{|E'|} q^{k(E')} \,, \label{eq:Zclust}
\end{equation}
where the sum is now over subsets of edges $E'$, with $|E'|$ representing the number of edges in $E'$ and $k(E')$ the number of connected components (including isolated vertices) in the subgraph $G' = (V,E')$. Note that the anti-ferromagnetic regime corresponding to real $K$ is now represented by $v \in [-1,0)$. Even so, the unphysical part of the phase diagram, $v<-1$, is of some mathematical interest \cite{Saleur91} and has been explored by various means \cite{SalasSokal2001,JacSal06,Jacobsen2012} for different lattices \cite{Jacobsen2012,Chang02,Chang04,JS12b} and boundary conditions \cite{Salas06,Salas07}.

In the ferromagnetic regime ($v>0$) the critical frontier is the locus of the transition between the high-temperature disordered (paramagnetic) phase and the low-temperature ordered (ferromagnetic) phase. Similarly, in the antiferromagnetic regime we expect a transition from the paramagnetic phase to a low-temperature phase of antiferromagnetic disorder. But in general, in the regime $v < 0$ the above studies \cite{Saleur91,SalasSokal2001,JacSal06,Jacobsen2012,Chang02,Chang04,JS12b,Salas06,Salas07} reveal a rich structure with several transitions between phases with various characteristics. We shall sometimes refer to the combined loci of these multiple transitions as the {\em critical manifold}.

From (\ref{eq:Zclust}), we can see that the limit $q \to 1$ is identical with bond percolation provided we set $v=p/(1-p)$, where $p$ is the probability an edge is open in the percolation model. This correspondence means not only that any critical point found in the Potts model solves the related percolation problem, but that results found in percolation can usually be generalized to arbitrary $q$ \cite{Wu2006,Ding2012}. For example, the ingenious argument used by Wierman to find the bond percolation threshold of the bow-tie lattice in 1984 \cite{Wierman1984} applies more or less directly to the Potts model, and this was done not long ago by Ding {\it et al.} \cite{Ding2012}.

In recent work \cite{Ziff2012}, it was shown that the percolation critical surface for the bow-tie lattice could be extended to include five different probabilities. This derivation is not completely rigorous, even from a physicist's perspective, as it partially depends on the use of the negative probability parameter regime of the triangular-lattice critical manifold. In fact, these parameters could be in the range $(-\infty,0]$ and thus were not really probabilities at all. Nevertheless, in the final formula all probabilities are positive and in the range $[0,1]$, and there is ample numerical \cite{Scullard2008} and analytical \cite{Scullard2010} evidence to suggest that the argument is correct. This leads to exact thresholds on many new lattices besides the simple bow-tie. 

\begin{figure}
\centering
\includegraphics{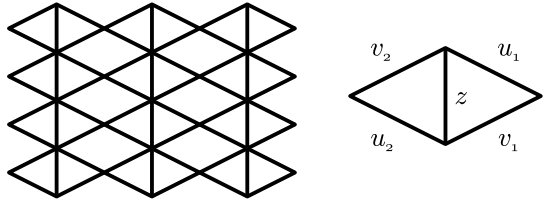}
\caption{The bow-tie lattice with five different edge-interactions.}
\label{fig:bowtie2}
\end{figure}

In this paper, we extend this result to the five-interaction Potts model (Figure \ref{fig:bowtie2}). The approach, laid out in the next section, is essentially identical to that used in \cite{Ziff2012}. In the Potts setting, the ``negative probability'' region corresponds to the anti-ferromagnetic regime and thus appears more natural, at least from a physical standpoint.

\begin{figure} [htbp]
\centering
\includegraphics{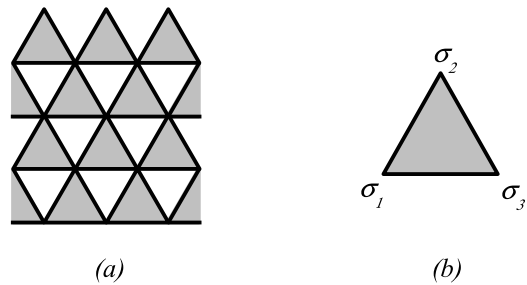}
\caption{a) Triangular-type hypergraph on which Potts critical frontiers can be found exactly; b) spins on the external vertices.}
\label{fig:triangular}
\end{figure}

\section{Exact solutions}
\begin{figure} [htbp]
\centering
\includegraphics{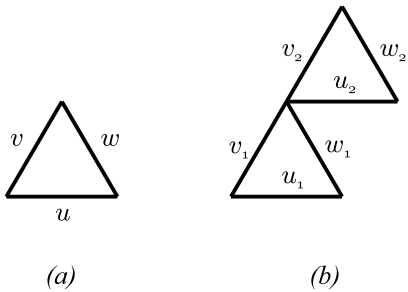}
\caption{a) Triangular lattice with three different Potts interactions. The critical frontier is given by equation (\ref{eq:tri}); b) generalization to six interactions. The critical manifold is unknown in general, but imposing the conditions (\ref{eq:tri1}) and (\ref{eq:tri2}) gives a critical system.}
\label{fig:triangles}
\end{figure}

One class of lattices with exactly known Potts critical manifolds is shown in Figure \ref{fig:triangular}a, where the shaded triangle stands for a generic configuration of spins and edges. In terms of the spins on the external vertices (Figure \ref{fig:triangular}b) this corresponds to effective interactions between groups of two and three spins. These interactions can be expanded in terms of clusters, generalizing \cite{WuLin1980} the construction of (\ref{eq:Zclust}). The critical point of this system is then given by \cite{Wu2006,WuLin1980}
\begin{equation}
qA-C=0 \ ,
\end{equation}
where $A$ is the weight of a triangle inside which none of the external spins are connected,
and $C$ is the weight for a triangle with all three spins connected.
Applying this formula to the simple triangular lattice, with the edge interactions shown in Figure~\ref{fig:triangles}a, we find $A=1$ and $C=u v w + u v + u w + v w$ leading to the critical frontier
\begin{equation}
q - u v w - u v - u w - v w =0 \ . \label{eq:tri}
\end{equation}
Removing the $w$-edge by setting $w=0$, we get the critical frontier for the square lattice
\begin{equation}
 q - u v =0 \ . \label{eq:sq}
\end{equation}
In equation (\ref{eq:tri}), we can single out an edge, say $w$, and write
\begin{equation}
 w = \frac{q-u v}{u v + u + v} \ . \label{eq:w}
\end{equation}
This form makes clear the content of the solution; the couplings $u$ and $v$ can be chosen arbitrarily, and then (\ref{eq:w}) provides the critical value of $w$. Note the presence of the square critical frontier, equation (\ref{eq:sq}), in the numerator of (\ref{eq:w}). The square lattice is super-critical if $uv>q$, and thus on the triangular lattice if we choose $uv>q$ the lattice is super-critical already and the role of the $w$-bond in equation (\ref{eq:w}) is to return the system to the critical point. This is accomplished by assigning $w$ an {\it anti}-ferromagnetic coupling, i.e., $-1<w<0$.

A generalization of the inhomogeneous triangular system is that shown in Figure \ref{fig:triangles}b, in which we have two types of triangles, $(u_1,v_1,w_1)$ and $(u_2,v_2,w_2)$. We can obtain a critical system by independently setting
\begin{equation}
 q - u_1 v_1 w_1 - u_1 v_1 - u_1 w_1 - v_1 w_1 =0 \label{eq:tri1}
\end{equation}
and
\begin{equation}
 q - u_2 v_2 w_2 - u_2 v_2 - u_2 w_2 - v_2 w_2 =0 \ . \label{eq:tri2}
\end{equation}
That this system is critical is a consequence of the difference property of quantum integrable systems (see the Appendix for the technical details). Of course, this is not the full critical frontier of the 6-interaction triangular lattice, which is unknown, rather these formulas occupy a sub-region of the full manifold.

\begin{figure}
\centering
\includegraphics{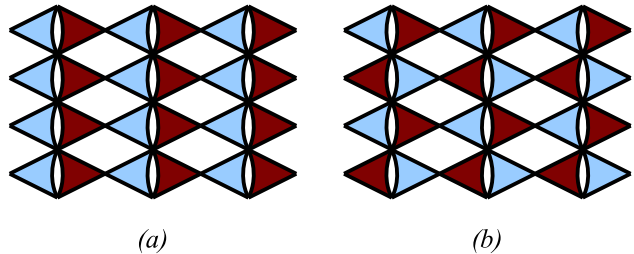}
\caption{A 3-uniform hypergraph on which Potts critical frontiers can be found exactly. Triangles of the same color have the same interactions on the edges.}
\label{fig:3uniform}
\end{figure}

\begin{figure}
\centering
\includegraphics{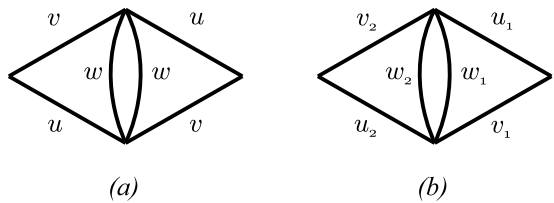}
\caption{a) Bow-tie lattice with three interactions for which the critical frontier is identical with that of the simple triangular lattice, equation (\ref{eq:tri}); b) generalization to six interactions.}
\label{fig:BTuvw}
\end{figure}

\section{Bow-tie lattice}
The critical manifold (\ref{eq:tri}) follows from a duality argument, the main ingredient of which is the fact that the unit cell of the triangular lattice lies between three boundary vertices as does that of its dual, the hexagonal lattice. In addition to lattices in which the triangular faces are more complicated \cite{Wu2006}, there are also alternative arrangements of triangles that have this property and thus their critical manifolds can be found exactly. An example can be seen in Figure \ref{fig:3uniform} (ignoring the colors for now). If the triangles are given the interactions $(u,v,w)$ as in Figure \ref{fig:BTuvw}a, the critical frontier for this lattice is given by (\ref{eq:tri}), i.e., identical to that of the ordinary triangular lattice. Now, the two $w$-edges in parallel may be combined into a single one, $z$, using the parallel reduction formula \cite{Sokal05},
\begin{equation}
 z=w^2+2w
\end{equation}
and we may substitute equation (\ref{eq:w}) for $w$ to get the Potts critical manifold for the bow-tie lattice, 
\begin{eqnarray}
& &q(q + 2 u + 2 v) - 2 u^2 v - 2 u v^2 - u^2 v^2 \cr
& &-z(u^2 + v^2 + 2 u v + 2 u^2 v + 2 u v^2 + u^2 v^2)\cr
& &=0 
\label{eq:bowtie3}
\end{eqnarray}
generalizing to arbitrary $q$ the well-known result for percolation \cite{Wierman1984}.

The formula (\ref{eq:bowtie3}) was derived in a recent work of Ding {\it et al.} \cite{Ding2012}. However, somewhat more is possible. Here, we show that the inhomogeneous interactions can be extended over neighboring triangles, giving a critical frontier involving all five interactions in a single formula. To do this, we note that, as for the ordinary triangular lattice, it is not necessary that all triangles be identical. The left- and right-pointing triangles may have different interactions, as indicated by the colors in Figure \ref{fig:3uniform}a with the assignments shown in Figure \ref{fig:BTuvw}b. As before, we can set the blue and red triangles separately to their critical values by requiring
\begin{equation}
w_1 = \frac{q-v_1 u_1}{u_1 v_1 + u_1 + v_1} \label{eq:w1}
\end{equation}
and
\begin{equation}
w_2 = \frac{q-v_2 u_2}{u_2 v_2 + u_2 + v_2}\ , \label{eq:w2}
\end{equation}
and the whole system is therefore at a critical point. However, we may now use the parallel reduction formula \cite{Sokal05} for the abutting $w$-interactions to combine them into the effective $z$ according to
\begin{equation}
z=w_1 w_2 + w_1 + w_2 \ . \label{eq:z}
\end{equation}
Now we simply substitute (\ref{eq:w1}) and (\ref{eq:w2}) into (\ref{eq:z}) and multiply away denominators to get
\begin{eqnarray}
& &q(q + u_1 + u_2 + v_1 + v_2) - u_1 u_2 v_1 v_2 \cr
& &-u_1 u_2 v_1 - u_1 u_2 v_2 - u_1 v_1 v_2 - u_2 v_1 v_2 \cr
& &-z(u_1 u_2 + u_2 v_1 + u_1 v_2 + v_1 v_2 + u_1 u_2 v_1 +\cr
& &u_1 u_2 v_2 + u_1 v_1 v_2 + u_2 v_1 v_2 + u_1 u_2 v_1 v_2) =0
\label{eq:bowtie5}
\end{eqnarray}
which is the critical manifold of the bow-tie system shown in Figure \ref{fig:bowtie2}. This reduces to the previously-obtained critical surface for percolation \cite{Scullard2010,Ziff2012} by setting $q=1$, but is a new result for the Potts model. It reduces to the formula (\ref{eq:bowtie3}) of Ding {\it et al.} in the special case $u_1=u_2=u$ and $v_1=v_2=v$, and to Wu's checkerboard formula \cite{Wu79} if $z=0$, providing another derivation of that result.

A word is in order about the meaning of the equation (\ref{eq:bowtie5}). Although we represent the effect of the two interactions $w_1$ and $w_2$ by a single interaction, $z$, both the $w$-bonds are really still in the problem. Consider the checkerboard case, $z=0$. Here, we choose $u_1$ and $v_1$ arbitrarily, and then $w_1$ must take its critical value (\ref{eq:w1}). Then, we select $w_2$ according to (\ref{eq:z}) in order to enforce $z=0$. We can now choose an arbitrary value for $u_2$, but $v_2$ must be determined from (\ref{eq:tri2}) to enforce criticality of the left triangle in Figure \ref{fig:BTuvw}b. The result is that we have chosen four interactions arbitrarily and were able to find the value of the fifth to give a critical system. This procedure is represented by the single formula (\ref{eq:bowtie5}). From this discussion, we can see that the statement ``$z=0$'' really corresponds to a family of values $(w_1,w_2)$ that exactly cancel this interaction, from which we must draw one member according to the values of the other interactions in the problem. This reasoning extends to any other value chosen for $z$.

\begin{figure}
\centering
\includegraphics{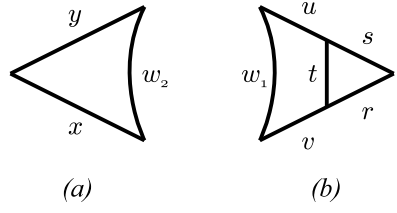}
\caption{Using a) for the blue (left-pointing) triangles in Figure \ref{fig:3uniform}a, and b) for the red triangles leads to the Potts critical frontier of the lattice in Figure \ref{fig:ASbowtie}a.}
\label{fig:ASBcons}
\end{figure}

\section{Asymmetric bow-tie lattices}
Aside from assigning different probabilities to the right and left sides, we may use more complicated triangular faces in place of the shaded red and blue triangles in Figure \ref{fig:3uniform}, and consequently, it is not necessary that the two sides contain the same graph. We may, for example, use the graph in Figure \ref{fig:ASBcons}a for the blue triangles in Figure \ref{fig:3uniform}a, and the graph in Figure \ref{fig:ASBcons}b, which is just the unit cell of the martini-A lattice \cite{Scullard2006,Ziff2006,Wu2006}, for the red. The critical frontier for simple triangle is given as before,
\begin{equation}
w_2=\frac{q-xy}{xy+x+y} \ . \label{eq:triw2}
\end{equation}
To find the critical point of Fig. \ref{fig:ASBcons}b, we employ the condition (\ref{eq:tri}) with $A$, the weight of configurations in which none of the boundary vertices are connected, given by
\begin{eqnarray}
A&=&q^2+q(r+s+t+u+v)+st+rt+rs\cr
&+&tu+tv+uv+ru+sv+rst
\end{eqnarray}
and $C$, the weight in which all three are connected, by
\begin{eqnarray}
C&=&r s t u v w_1 + r s t u v + r s t u w_1 + r s t v w_1 \cr
&+& r s u v w_1 + r t u v w_1 + s t u v w_1 + r s u v \cr
&+& r t v w_1 + r s v w_1 + r u v w_1 + r s u w_1 \cr
&+& s t u w_1 + r u v w_1 + r t u v + s t u v \cr
&+& s t v w_1 + r t u w_1 + q (r v w_1 +  s u w_1),
\end{eqnarray}
with the result
\begin{equation}
w_1=\frac{\alpha}{\beta} \label{eq:Abarw1}
\end{equation}
where
\begin{eqnarray}
\alpha &\equiv& q^3 + q^2 (r + s + t + u + v) \cr
&+& q (r s + r t + s t + r u + t u + s v + t v + u v) \cr
&+& q r s t - r s u v - r t u v - s t u v - r s t u v
\end{eqnarray}
and
\begin{eqnarray}
\beta &\equiv&q (s u + r v) + r s u + r t u + s t u + r s v \cr
&+& r t v + s t v + r u v  + s u v + r s t v + r s t u \cr
&+& r s u v + r t u v + s t u v + r s t u v \ . 
\end{eqnarray}
Now we plug (\ref{eq:triw2}) and (\ref{eq:Abarw1}) into (\ref{eq:z}) and multiply by $\beta(x+y+xy)$ to get the full 8-interaction critical frontier. Setting all interactions equal gives the homogeneous critical frontier displayed in Table \ref{tab:ASbowtie}a.

\begin{figure} [htbp]
\centering
\includegraphics{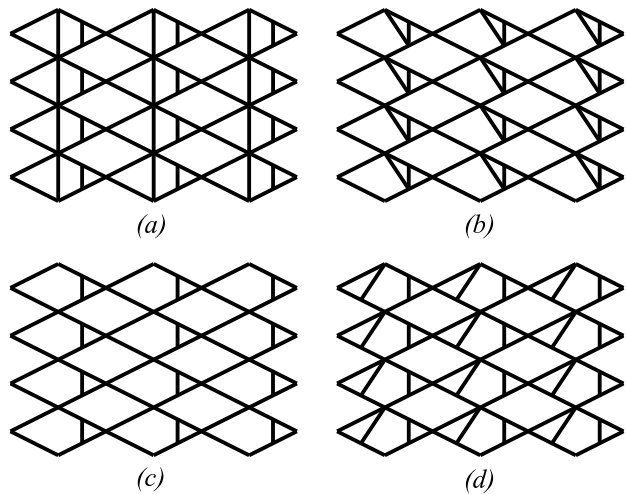}
\caption{Asymmetric bow-tie lattices. The critical frontiers are given in Table \ref{tab:ASbowtie}.}
\label{fig:ASbowtie}
\end{figure}

Many constructions of this type are possible and we give a very small sample in the rest of Figure~\ref{fig:ASbowtie}, with their homogeneous critical frontiers also reported in Table~\ref{tab:ASbowtie}. The inhomogeneous frontiers for all the lattices of Figure~\ref{fig:ASbowtie} are included in the supplemental material to this submission in the file {\tt SJ12\_EPL.m}, which can be processed in {\sc Mathematica} or opened with an ordinary text editor.

\begin{figure} [htbp]
\centering
\includegraphics{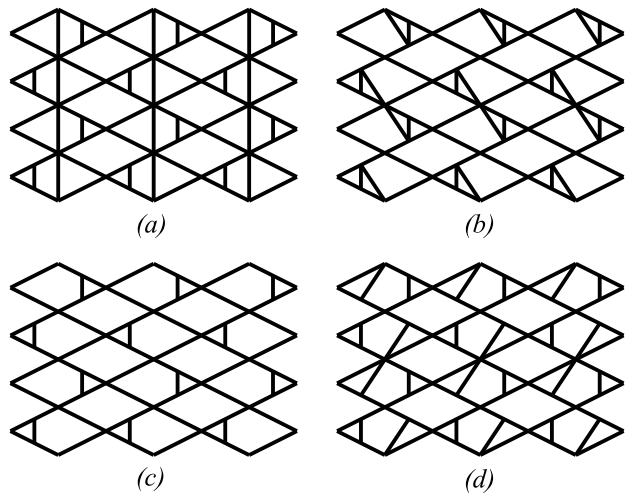}
\caption{Asymmetric bow-tie lattices with the same critical frontiers as those in Figure \ref{fig:ASbowtie}.}
\label{fig:ASbowtie2}
\end{figure}

The argument that leads to these results applies equally well to the configuration in Figure \ref{fig:3uniform}b, corresponding to rotating every second row by 180 degrees. Again using the graphs of Figures \ref{fig:ASBcons}a and b for the blue and red triangles gives the lattice in Figure \ref{fig:ASbowtie2}a. The other lattices in Fig. \ref{fig:ASbowtie2} are similarly obtained from those in \ref{fig:ASbowtie} and have the corresponding critical frontiers.

\begin{table}
\begin{center}
\begin{tabular}{c|l}
 lattice & \ \ \ \ \ \ \ \ \ \ frontier \\
 \hline \hline
  &  \\
a & $q (q^3 + 7 q^2 v + 20 q v^2 + 24 v^3 + q v^3)$  \\
  & $-2 q v^4 - 30 v^5 - 2 q v^5 - 25 v^6 - 8 v^7 - v^8$ \\
\hline
  &  \\
b & $q (q^3 + 8 q^2 v + 26 q v^2 + 35 v^3 + 2 q v^3 + 5 v^4)$  \\
  & $-29 v^5 - 24 v^6 - 8 v^7 - v^8$  \\
\hline
  &  \\
c & $q (q^3 + 7 q^2 v + 20 q v^2 + 24 v^3 + q v^3 + 2 v^4)$  \\
  & $-14 v^5 - 7 v^6 - v^7$\\
\hline
  &  \\
d & $q (q^4 + 9 q^3 v + 35 q^2 v^2 + 71 q v^3 + 2 q^2 v^3$ \\
  & $+ 64 v^4 + 11 q v^4 + 10 v^5)$  \\
  & $-39 v^6 - 2 q v^6 - 31 v^7 - 9 v^8 - v^9$\\
\end{tabular}
\caption{Critical frontiers for the lattices shown in Figures \ref{fig:ASbowtie} and \ref{fig:ASbowtie2}.}
\label{tab:ASbowtie}
\end{center}
\end{table}

\section{Discussion}

We have presented new exact critical frontiers for the Potts model on certain two-dimensional lattices. These new solutions employ the special structure of the bow-tie lattice, enabling connection of neighboring triangles and leading to exact solutions on lattices that are realizations of 4-uniform hypergraphs. 

Note that these lattices are not of the typical exactly solvable type in which the unit cell is contained within three boundary vertices, or terminals; that is, they are not realizations of 3-uniform hypergraphs. Rather, these are {\it four}-terminal lattices. However, they are a very special subset of this class in that they can be derived from three-terminal lattices by fusing two edges. This fact leads to the related observation that the duals of these lattices are also four-terminal lattices, a property not generally shared by such graphs. Our method is thus not a recipe for finding the critical frontiers of general four-uniform hypergraphs, such as the kagome lattice. Its dual, the diced lattice, is not of the four-terminal type.

We finally mention that we have verified our results by computing the critical polynomials \cite{Jacobsen2012} corresponding to the lattices of Figure~\ref{fig:ASbowtie}, using square bases of size $2 \times 2$, and Figure \ref{fig:ASbowtie2} using rectangular bases of size $1 \times 2$ unit cells. We found that for each of these lattices, the critical polynomial contains a factor which coincides with the frontier given in Table~\ref{tab:ASbowtie}. This is in line with the observations of \cite{Jacobsen2012,JS12b} that exact solvability implies that the smallest possible basis ($1 \times 1$) produces the exact critical frontier, and that larger bases (such as $2 \times 2$) factorize, shedding a factor which is equal to the exact critical frontier.
\begin{figure} [tbp]
\centering
\includegraphics{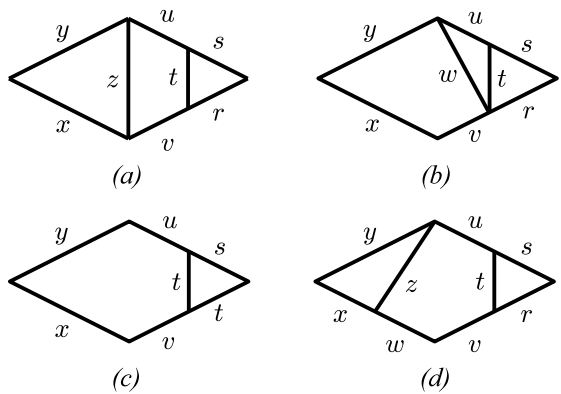}
\caption{Potts interactions on the lattices in Figures~\ref{fig:ASbowtie} and \ref{fig:ASbowtie2}.}
\label{fig:ASBTassgn}
\end{figure}
\section*{Appendix: Relation to quantum integrability}
\label{sec:appA}

\begin{figure}
\centering
\includegraphics{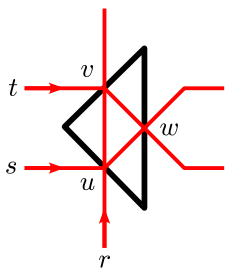}
\caption{Part of the medial graph ${\cal M}(G)$ corresponding to a left-pointing triangle of $G$.}
\label{fig:medial}
\end{figure}

It is well-known \cite{BaxterKellandWu1976} that the Potts model on any planar graph $G$ is equivalent to a loop model on the associated medial graph ${\cal M}(G)$.
The portion of the medial graph corresponding to the leftmost triangle in Figure \ref{fig:triangles}a is shown in Fig.~\ref{fig:medial}.


The integrable R-matrix of the loop model takes the form \cite{Baxter_book}
\begin{equation}
 \check{R}_i(\theta) = \sin(\gamma-\theta) {\sf I}_i + \sin(\theta) {\sf E}_i \,.
 \label{Rcheck}
\end{equation}
We interpret the coefficients as the set of Boltzmann weights associated with a vertex $i$ of ${\cal M}(G)$ (i.e., with an edge $e$ of $G$). Here $\theta$ is the difference of the two
spectral parameters carried by the oriented lines of ${\cal M}(G)$ that intersect at $i$. The Temperley-Lieb generator ${\sf E}_i$ can be interpreted
graphically as the contraction of the two lines going into $i$, and the identity operator ${\sf I}_i$ is the reflexion of those lines at $i$.
The equivalence between the Potts model and the loop model implies that the weight of a loop is $n = \sqrt{q}$, and we have parameterised
this as $n = 2 \cos(\gamma)$ in (\ref{Rcheck}). Moreover, the edge weights $u$ get traded for new weights $x = u q^{-1/2}$ that
are ratios of those appearing in front of the terms in (\ref{Rcheck}). More precisely, we have $x = \sin(\theta) / \sin(\gamma-\theta)$ if the two
oriented lines of ${\cal M}(G)$ go into the edge $e$ from the same side, and $x = \sin(\gamma-\theta) / \sin(\theta)$ if they go into $e$ from opposite sides.

Returning to Figure~\ref{fig:medial} we note that integrability requires that the spectral parameters $r$, $s$ and $t$ follow the lines of ${\cal M}(G)$ through
the intersection at vertices. In particular, on the right side of the figure the $s$-line is on the top and the $t$-line is on the bottom. But since in the bow-tie
lattice the left-pointing triangle is followed by a right-pointing one (see Figure~\ref{fig:triangles}), the two lines get permuted back after traversing a complete bow-tie motif. In particular,
the assignation of spectral parameters to each ``horizontal'' line is compatible with periodic boundary conditions.

For a choice $(r,s,t)$ of spectral parameters, the edge weights in the Potts model are therefore
\begin{eqnarray}
 u &=& q^{1/2} \, \frac{\sin(s-r)}{\sin(\gamma-s+r)} \,, \nonumber \\
 v &=& q^{1/2} \, \frac{\sin(\gamma-t+r)}{\sin(t-r)} \,, \nonumber \\
 w &=& q^{1/2} \, \frac{\sin(t-s)}{\sin(\gamma-t+s)} \,.
\end{eqnarray}
It follows that the quantity $u v w + u v + u w + v w - q$ appearing on the left-hand side of (\ref{eq:tri}) is proportional to
\begin{eqnarray}
 & & 2 \cos(\gamma) \sin(s-r) \sin(\gamma-t+r) \sin(t-s) \nonumber \\
 &+& \sin(s-r) \sin(\gamma-t+r) \sin(\gamma-t+s) \nonumber \\
 &+& \sin(s-r) \sin(t-s) \sin(t-r) \nonumber \\
 &+& \sin(\gamma-t+r) \sin(t-s) \sin(\gamma-s+r) \nonumber \\
 &-& \sin(\gamma-s+r) \sin(t-r) \sin(\gamma-t+s).
\end{eqnarray}
But this is identically zero, by a trigonometric identity. The condition (\ref{eq:tri}) is therefore simply equivalent to the integrability of the Potts model.
The same reasoning applies of course to the right-pointing triangles, after some relabeling of the parameters.
In particular we infer that transfer matrices $T(s,t | r)$ and $T(s',t' | r)$ with different choices of the horizontal spectral parameters commute,
and hence the transfer matrix commutes with an infinite set of conserved quantities. It follows that the bow-tie lattice Potts model satisfying (\ref{eq:tri1})--(\ref{eq:tri2})
is critical.

Summarizing, the appearance of the constraint (\ref{eq:tri}) between the edge weights $(u,v,w)$ is due to the fact that the three spectral parameters $(r,s,t)$
admit only two independent differences $r-s$ and $s-t$ (since $r-t = (r-s) + (s-t)$), and the requirement that the integrable R-matrix $\check{R}(\theta)$ satisfies
the difference property (i.e., its argument $\theta$ is the difference of two spectral parameters).

\section*{Acknowledgments}
CRS thanks Robert Ziff, John Wierman and Matthew Sedlock for the collaboration that led to this work.
The work of JLJ was supported by the Agence Nationale de la Recherche
(grant ANR-10-BLAN-0414:~DIME) and the Institut Universitaire de France.

\section*{References}
\bibliography{SJ12c}






\end{document}